\documentclass[12pt,a4paper]{article}
\usepackage{amsmath,caption,amsfonts,setspace,mathrsfs,mathtools}
\usepackage[top=1.4in, bottom=1.4in, left=1in, right=1in]{geometry}

\usepackage[round]{natbib}
\usepackage{color,soul}

\DeclareCaptionStyle{italic}[justification=centering]{labelfont={bf},textfont={it},labelsep=colon}
\usepackage{graphicx,psfrag,epsf,textcomp}
\usepackage{enumerate, dsfont}
\usepackage{natbib}
\usepackage{url,xcolor} 
\usepackage{booktabs, subfig, bm, paralist,mathpazo,tikz,todonotes,longtable,microtype}

\usepackage[pdftex,colorlinks=true]{hyperref}
\definecolor{darkblue}{rgb}{0,0,.6}
\hypersetup{citecolor=darkblue,linkcolor=darkblue,urlcolor=darkblue}
\newcommand{\argmax}{\operatornamewithlimits{argmax}}

\newcommand{\blind}{0}

\graphicspath{{plots/}}
\DeclareMathOperator*{\argmin}{\arg\!\min}
\newsavebox\CBox
\def\textBF#1{\sbox\CBox{#1}\resizebox{\wd\CBox}{\ht\CBox}{\textbf{#1}}}

\begin{document}

\def\spacingset#1{\renewcommand{\baselinestretch}{#1}\small\normalsize} \spacingset{1}

\if0\blind
{
  \title{\bf A robust functional time series forecasting method}
  \author{
    Han Lin Shang\footnote{Postal address: Research School of Finance, Actuarial Studies and Statistics, Level 4, Building 26C, Australian National University, Acton Canberra, ACT 2601, Australia; Telephone: +61(2) 612 50535; Fax: +61(2) 612 50087; Email: hanlin.shang@anu.edu.au}\\
 Australian National University 
 }
  \maketitle
} \fi

\if1\blind
{
\title{\bf A robust functional time series forecasting method}
\maketitle
} \fi

\bigskip
\begin{abstract}
Univariate time series often take the form of a collection of curves observed sequentially over time. Examples of these include hourly ground-level ozone concentration curves. These curves can be viewed as a time series of functions observed at equally spaced intervals over a dense grid. Since functional time series may contain various types of outliers, we introduce a robust functional time series forecasting method to down-weigh the influence of outliers in forecasting. Through a robust principal component analysis based on projection pursuit, a time series of functions can be decomposed into a set of robust dynamic functional principal components and their associated scores. Conditioning on the estimated functional principal components, the crux of the curve-forecasting problem lies in modeling and forecasting principal component scores, through a robust vector autoregressive forecasting method. Via a simulation study and an empirical study on forecasting ground-level ozone concentration, the robust method demonstrates the superior forecast accuracy that dynamic functional principal component regression entails. The robust method also shows the superior estimation accuracy of the parameters in the vector autoregressive models for modeling and forecasting principal component scores, and thus improves curve forecast accuracy.

\vspace{.2in}
\end{abstract}

\newpage

\def\spacingset#1{\renewcommand{\baselinestretch}{#1}\small\normalsize} \spacingset{1}
\spacingset{1.45}

\section{Introduction}

Recent advances in computer technology and its wide application in various disciplines has enabled the collection, storage, and analysis of data of huge sizes, large dimensions, and various formats. To analyze these data, functional data analysis has received increasing attention in theoretical and applied research. As a stream of functional data, a functional time series often consists of random functions observed at regular time intervals. Depending on whether or not the continuum is also a time variable, functional time series can arise by separating a continuous time record into natural consecutive intervals such as days, months or years. Conversely, functional time series can also arise when the observations in a time period can be considered finite realizations of an underlying continuous function such as age-specific demographic rates. In either case, the functions obtained form a time series $\{\mathcal{X}_i, i\in \mathbb{Z}\}$, where each $\mathcal{X}_i$ is a random function $\mathcal{X}_i(t)$ and $t\in \mathcal{I}\subset R$ denotes a continuum bound within a finite interval. We refer to such data structures as functional time series.

\cite{HU07} and \cite{ANH15} use static or dynamic functional principal component analysis (FPCA) to decompose a functional time series into a set of functional principal components and their associated scores. The temporal dependence in the original functional time series is inherited by the correlation within each set of principal component scores and the possible cross-correlation between sets of principal component scores. Conditioning on the estimated basis functions, the forecast functions can be obtained by accurately modeling and forecasting the principal component scores. \cite{HU07} applied univariate time series forecasting techniques to forecast each set of principal component scores individually, while \cite{ANH15} considered multivariate time series forecasting methods, such as vector autoregressive (VAR) models, to capture any correlations between stationary principal component scores. Both univariate and multivariate time series forecasting methods have their advantages and disadvantages \citep[see][for more details]{PS07}.

The majority of statistical modeling and forecasting techniques used in the functional time series analysis so far assume that the dataset is free of outliers, despite the fact that outliers occur very frequently in functional time series \citep[e.g.][]{RAV15}. Outliers can be categorized into additive and innovative outliers. An observation is an additive outlier if only its value has been affected by contamination, while an outlier is an innovation outlier if the error term in a time series model is contaminated \citep[see][]{MMY06}.

Outlier analysis of functional time series comprises of two key issues: 
\begin{inparaenum}
\item[(1)] searching for the location and type of outliers in a contaminated functional time series (known as the ``outlier detection/diagnostic") \citep[e.g.][]{HS10, SG11}; 
\item[(2)] obtaining better estimates of parameters in the underlying functional time series model through the incorporation of outlier effects within a model (known as the ``outlier adjustment") \citep[e.g.][]{HU07, MY13}. Both outlier detection and adjustment are critical to model estimation and forecasting. 
\end{inparaenum}

We aim to contribute to the literature of functional time series and outlier adjustment by introducing a robust functional time series method by dynamic functional principal component regression. The method can model historical time series and produce robust forecasts by down-weighing the influence of possible outliers. The essence of this method is to obtain robust and reliable forecasts of multivariate principal component scores. To this end, we use the multivariate least trimmed squares (MLTS) estimator of \cite{ACV08} and \cite{CJ08} in a VAR model. This estimator is defined by minimizing a trimmed sum of the squared Mahalanobis distance, and it provides a robust and invariant estimator for the covariance matrix of the residuals, which can then be used for model selection based on a robust information criterion, such as the Bayesian information criterion (BIC). 

The outline of this paper is as follows. A functional time series forecasting method is described in Section~\ref{sec:2}. In Section~\ref{sec:3}, we introduce a robust functional time series forecasting method. The robustness of the VAR estimator is studied by a series of simulation studies in Section~\ref{sec:4}. Illustrated by an hourly ground-level ozone concentration data in Section~\ref{sec:5}, the robust and standard VAR models used to form an essential part of functional principal component regression are investigated and compared in terms of one-step-ahead point forecast accuracy. Section~\ref{sec:7} concludes the paper and outlines some ideas on how the robust functional time series forecasting method presented here can be further extended.

\section{A functional time series forecasting method}\label{sec:2}

We introduce a novel method for forecasting functional time series when data are free of outliers. The method relies on dynamic functional principal components and their scores extracted from the estimated long-run covariance function.

\subsection{Notation}

Let $\{\mathcal{X}_i, i\in \mathbb{Z}\}$ be an arbitrary functional time series. It is assumed that the observations $\mathcal{X}_i$ are elements of the square-integrable space $\mathcal{L}^2(\mathcal{I})$ equipped with the inner product $\langle x,y\rangle = \int_{\mathcal{I}} x(t)y(t)dt$, where $t$ symbolizes a continuum and $\mathcal{I}$ represents the function support range. Each function is a square-integrable function satisfying $\|\mathcal{X}_i\|^2 = \int_{\mathcal{I}}\mathcal{X}_i^2(t)dt <\infty$ and associated norm. All random functions are defined on a common probability space $(\Omega, A, P)$. The notation $\mathcal{X}\in L_{H}^p(\Omega, A, P)$ is used to indicate $\mathbb{E}(\|\mathcal{X}\|^p)<\infty$ for some $p>0$. When $p=1$, $\mathcal{X}(t)$ has the mean curve $\mu(t) = \mathbb{E}[\mathcal{X}(t)]$; when $p=2$, a non-negative definite long-run covariance function is given by
\begin{align}
\gamma_{\vartheta}(t,s) &= \text{Cov}[\mathcal{X}_0(s), \mathcal{X}_{\vartheta}(t)],  \notag\\
c_{\mathcal{X}}(t,s) &= \sum^{\infty}_{\vartheta = -\infty}\gamma_{\vartheta}(t,s), \qquad \forall s,t\in \mathcal{I},\label{eq:long_run}
\end{align}
where $\vartheta$ denotes a lag parameter. The long-run covariance function is a well-defined element in $\mathcal{L}^2(\mathcal{I}^2)$ under mild weak dependence and moment conditions. 

When the temporal dependence is weak, the long-run covariance function in~\eqref{eq:long_run} can be reasonably approximated by variance function alone. When the temporal dependence is moderate or strong, the long-run covariance function also encompasses the autocovariance function. 

Among many estimators for estimating the long-run covariance function, the commonly used approach is the kernel sandwich estimator of \cite{Andrews91} and \cite{AM92} in univariate time series. Recently, this estimator has been extended to functional time series by \cite{RS17}. 

\subsection{Estimation of long-run covariance}\label{sec:long_run_covariance}

It is of interest in many applied settings to estimate $c_{\mathcal{X}}(t,s)$ from a finite collection of samples $\left\{\mathcal{X}_1, \mathcal{X}_2, \dots,\mathcal{X}_n\right\}$. Given its definition as a bi-infinite sum, a natural estimator of $c_{\mathcal{X}}(t,s)$ is
\begin{align}\label{est-1}
\widehat{c}_{h,q}(t,s)=\sum_{\vartheta=-\infty}^{\infty}W_{q} \left( \frac{\vartheta}{h} \right) \widehat{\gamma}_{\vartheta}(t,s),
\end{align}
where $h$ is called the bandwidth parameter,
\begin{align*}
   \widehat{\gamma}_{\vartheta}(t,s)=\left\{
     \begin{array}{lr}
      \displaystyle \frac{1}{n}\sum_{j=1}^{n-\vartheta}\left[\mathcal{X}_j(t)-\overline{\mathcal{X}}(t)\right]\left[\mathcal{X}_{j+\vartheta}(s)-\overline{\mathcal{X}}(s)\right],\quad &\vartheta \ge 0
      \vspace{.3cm} \\
     \displaystyle \frac{1}{n}\sum_{j=1-\vartheta}^{n}\left[\mathcal{X}_j(t)-\overline{\mathcal{X}}(t)\right]\left[\mathcal{X}_{j+\vartheta}(s)-\overline{\mathcal{X}}(s)\right],\quad &\vartheta < 0,
     \end{array}
   \right.
\end{align*}
is an estimator of $\gamma_{\vartheta}(t,s)$, and $W_q$ is a symmetric weight function with bounded support of order $q$, which is to say that
\begin{align*}
&W_{q}(0)=1,\; W_q(u)\le 1, \;W_{q}(u)=W_{q}(-u),\;W_{q}(u)=0\;\;\mbox{if}\;\; |u|>m \;\;\mbox{for some } m>0,\\
& \mbox{ and }W_{q} \mbox{ is } \mbox{continuous on}\;\;[-m,m], \notag
\end{align*}
and there exists $w$ satisfying
\begin{align*}
0< w = \lim_{x\to 0} |x|^{-q}\left[1-W_{q}(x)\right] < \infty.
\end{align*}

The estimator in~\eqref{est-1} was introduced in~\cite{HKR13}, further studied in \cite{RS17} on the optimal selection of bandwidth in finite samples. 

\subsection{Dynamic functional principal component analysis}\label{sec:21}

Via right integration, $c_{\mathcal{X}}(t,s)$ defines a Hilbert-Schmidt integral operator on $\mathcal{L}^2(\mathcal{I})$ given by
\begin{equation*}
\mathcal{K}_{\mathcal{X}}(\phi)(s) = \int_{\mathcal{I}}c_{\mathcal{X}}(t,s)\phi(t)dt,
\end{equation*}
whose eigenvalues and eigenfunctions are related to the dynamic functional principal components defined in \cite{HKH15}, and provide asymptotically optimal finite dimensional representations of the sample mean of dependent functional data.

Via Mercer's lemma, there exists an orthonormal sequence $(\phi_k)$ of continuous function in $\mathcal{L}^2(\mathcal{I})$ and a non-increasing sequence $(\lambda_1,\lambda_2,\dots)$ of positive number, such that
\begin{equation*}
c_{\mathcal{X}}(t,s) = \sum^{\infty}_{k=1}\lambda_k\phi_k(s)\phi_k(t),\qquad s,t \in \mathcal{I},
\end{equation*}
where $\lambda_k$ denotes the $k^{\text{th}}$ eigenvalue.

From the estimated long-run covariance in~\eqref{est-1}, a time series of functions $\bm{\mathcal{X}}(t) = \{\mathcal{X}_1(t),\dots,\mathcal{X}_n(t)\}$ can be decomposed into orthogonal functional principal components and their associated principal component scores, given by
\begin{align*}
\mathcal{X}_i(t) &= \mu(t) + \sum^{\infty}_{k=1}\beta_{i,k}\phi_k(t) \\
&=\mu(t) + \sum^K_{k=1}\beta_{i,k}\phi_k(t) + e_i(t), \qquad i=1,\dots,n,
\end{align*}
where $\mu(t)$ denotes the mean function, $\{\phi_1(t),\dots,\phi_K(t)\}$ denotes a set of first $K$ functional principal components, $\bm{\beta}_1 = (\beta_{1,1},\dots,\beta_{1,n})^{\top}$ and $\{\bm{\beta}_1,\dots,\bm{\beta}_K\}$ denotes a set of principal component scores where $\bm{\beta}_k\sim  N(0,\lambda_k)$ and $^{\top}$ denotes matrix transpose, $e_i(t)$ denotes the model truncation error function with mean zero and finite variance, and $K<n$ denotes the number of retained functional principal components.

There are several approaches for selecting $K$, and we consider predictive cross validation (CV) using holdout curves \citep{RS91}. The value of $K$ is chosen as the minimum that produces the smallest averaged predictive error in a validation data set of length $n_1$, defined by
\begin{equation*}
K=\argmin_{K:K\geq 1}\frac{1}{\mathfrak{p}\times n_1}\sum_{w=1}^{\mathfrak{p}}\sum_{\upsilon=1}^{n_1}\left[\mathcal{X}_{\upsilon}(t_w) - \widehat{\mathcal{X}}_{\upsilon}(t_w)\right]^2,
\end{equation*}
where $\widehat{\mathcal{X}}_{\upsilon}(t_w)$ represents the forecast at the $w^{\text{th}}$ grid point of the $\upsilon^{\text{th}}$ sample in the validation data set, and $\mathfrak{p}$ denotes the number of grid points.

\subsection{Vector autoregressive models}

The principal component scores can be modeled and forecasted via multivariate time series methods. These principal component scores are orthogonal to each other, but correlations may exist at various lags between two sets of principal component scores. For such multiple stationary time series, it has become very popular to use VAR due to its ability to model the temporal dependence within and between variables (i.e. principal component scores). The widespread use of VAR is largely due to the fact that it can be expressed as a multivariate linear regression. Let $\bm{\beta}_{\mathfrak{i}}$ be a $K$-dimensional stationary multivariate time series. A VAR model of order $\omega$, denoted by VAR$(\omega)$, is given by
\begin{equation*}
\bm{\beta}_{\mathfrak{i}} = \bm{B}_0 + \bm{B}_1\bm{\beta}_{\mathfrak{i}-1} + \cdots + \bm{B}_{\omega}\bm{\beta}_{\mathfrak{i}-\omega} + \bm{\varepsilon}_{\mathfrak{i}}, \qquad \mathfrak{i} = \omega+1,\dots,n, 
\end{equation*}
where $\bm{B}_1,\dots,\bm{B}_{\omega} (\bm{B}_{\omega}\neq 0)$ are $(K\times K)$ unrestricted coefficient matrices, and $\bm{B}_0$ is a fixed $(K\times 1)$ vector of intercept terms, and $\bm{\varepsilon}_{\mathfrak{i}}$ is a $K$-dimensional white noise with covariance matrix $\bm{\Sigma}$ that is assumed to nonsingular. The $K$-dimensional error terms are supposed to be independent and identically distributed (iid) with a density of 
\begin{equation}
f_{\bm{\Sigma}}(\bm{u}) = \frac{g(\bm{u}^{\top}\bm{\Sigma}^{-1} \bm{u})}{\sqrt{\text{det}(\bm{\Sigma})}}, \label{eq:error_den}
\end{equation}
where $f(\cdot)$ denotes a probability density function, $\bm{u}$ denotes the mean of the error term, $\bm{\Sigma}$ denotes a positive definite matrix of the error term, and $g$ denotes a positive function that is assumed to have a strictly negative derivative $g^{'}$ such that cumulative distribution function $F_{\bm{\Sigma}}$ is a unimodal elliptically symmetric distribution around the origin. Via a multivariate linear regression model, the VAR$(\omega)$ can be re-written as VAR(1),
\begin{equation*}
\bm{\beta}_{\mathfrak{i}} = \bm{B}^{\top}\bm{x}_{\mathfrak{i}} + \bm{\varepsilon}_{\mathfrak{i}}, \qquad \mathfrak{i}=\omega+1,\dots,n 
\end{equation*}
with $\bm{x}_{\mathfrak{i}} = \left(\bm{1}, \bm{\beta}_{\mathfrak{i}-1}^{\top}, \dots, \bm{\beta}_{\mathfrak{i}-\omega}^{\top}\right)^{\top}\in R^q$ and $q=K\omega+1$. The matrix $\bm{B} = \left(\bm{B}_0^{\top}, \bm{B}_1^{\top}, \dots, \bm{B}_{\omega}^{\top}\right)^{\top}\in R^{q\times K}$ contains all unknown regression coefficients. Let $\bm{X} = (\bm{x}_{\omega+1}, \dots, \bm{x}_n)^{\top}\in R^{(n-\omega)\times q}$ denote the matrix containing the values of the explanatory variables and $\bm{Y} = (\bm{\beta}_{\omega+1},\dots,\bm{\beta}_n)^{\top}\in R^{(n-\omega)\times K}$ be the matrix of responses, then we can express
\begin{equation*}
\bm{Y} = \bm{X}\bm{B} + \bm{A},
\end{equation*}
where $\bm{A}$ denotes $(n-\omega)\times K$ error matrix with mean zero. The unknown regression coefficients can then be estimated by the ordinary least squares (OLS) estimator, although it is not the optimal estimator. The OLS estimator is defined as
\begin{equation}
\bm{\widehat{B}}_{\text{OLS}} = \left(\bm{X}^{\top}\bm{X}\right)^{-1}\bm{X}^{\top}\bm{Y}, \label{eq:B_OLS}
\end{equation}
and it does not depend on $\bm{\Sigma}$. The error matrix $\bm{\Sigma}$ can be estimated by
\begin{equation}
\bm{\widehat{\Sigma}}_{\text{OLS}} = \frac{\left(\bm{Y} - \bm{X}\widehat{\bm{B}}_{\text{OLS}}\right)^{\top}\left(\bm{Y} - \bm{X}\widehat{\bm{B}}_{\text{OLS}}\right)}{n-(K+1)\omega-1}=\frac{\widehat{\bm{A}}^{\top}\widehat{\bm{A}}}{n-(K+1)\omega-1}, \label{eq:Var_OLS}
\end{equation}
where $\widehat{\bm{A}}$ denotes residual matrix, and the denominator is $\left[n-(K+1)\omega-1\right]$, which is the effective sample size \citep[see also][]{IV05}.

Conditioning on the estimated functional principal components and past curves, the $h$-step-ahead forecast principal component scores and forecast curves are given respectively by
\begin{align*}
\widehat{\bm{\beta}}_{n+h} &= \bm{x}_{n+h}^{\top} \times \widehat{\bm{B}}_{\text{OLS}}  \\
\widehat{\mathcal{X}}_{n+h}(t) &= \widehat{\bm{\beta}}_{n+h}\times \widehat{\bm{\phi}}(t),
\end{align*}
where $\widehat{\bm{\beta}}_{n+h}=\big(\widehat{\beta}_{n+h,1},\dots,\widehat{\beta}_{n+h,K}\big)$ denotes $h$-step-ahead forecast principal component scores, $\bm{x}_{n+h} = \left(\bm{1}, \bm{\beta}_{n+h-1}^{\top},\dots,\bm{\beta}_{n+h-\omega}^{\top}\right)^{\top}$ denotes lagged principal component scores, and $\widehat{\bm{\phi}}(t)$=$\big[\widehat{\phi}_1(t), \dots, \widehat{\phi}_K(t)\big]^{\top}$ denotes the estimated functional principal components.

\section{A robust functional time series forecasting method}\label{sec:3}

We introduce our proposed methods for forecasting functional time series when data contain outliers. Our methods begin by removing outlying functions, then extract dynamic functional principal components and forecast their scores via a robust VAR procedure with a robust BIC criterion.

\subsection{Robust functional principal component analysis}\label{sec:3.1}

The presence of outliers can seriously affect the estimators of mean and variance for a functional time series, and can have a substantial influence on functional principal component decomposition. To down-weigh the effect of outliers, we use a robust FPCA considered previously in \cite{HU07}. Based on the projection-pursuit approach, the FPCA utilizes reflection-based algorithm for principal component analysis (RAPCA) \citep{HRV02}. The algorithm takes the first quartile of pairwise score differences as the measure of dispersion, thus
\begin{equation*}
S(\beta_{1,k},\dots,\beta_{n,k}) = 2.2219\times c_n\times \{|\beta_{i,k} - \beta_{j,k}|; i<j\}_{(\tau)},
\end{equation*}
where $S(\cdot)$ denotes a measure of dispersion, such that the sample variance that can be computed easily; $\tau={\lfloor n/2 \rfloor + 1 \choose 2}$ where $\lfloor n/2 \rfloor$ is the largest integer less than or equal to $n/2$; $c_n$ is a small-sample correction factor to make dispersion function $S(\cdot)$ unbiased, and the value of $c_n\rightarrow 1$ for increasing sample size $n$. In Table~\ref{tab:cn_val}, we list a number of $c_n$ values \citep[see also][]{RC93}. 
\begin{table}[!htbp]
\tabcolsep 0.45in
\centering
\caption{List of correction factors $c_n$.}\label{tab:cn_val}
\begin{singlespace}
\begin{tabular}{@{}ll@{}}
\toprule
$n$ & $c_n$ \\\midrule
2 & 0.399 \\
3 & 0.994 \\
4 & 0.512 \\
5 & 0.844 \\
6 & 0.611 \\
7 & 0.857 \\
8 & 0.669 \\
9 & 0.872 \\
$n$ \text{mod} 2 ==1 & $n/(n+1.4)$ \\
$n$ \text{mod} 2 ==0 & $n/(n+3.8)$ \\\bottomrule
\end{tabular}
\end{singlespace}
\end{table}

Using the RAPCA algorithm, we can obtain initial estimates of $\beta_{i,k}$ and $\phi_k(t)$ for $k=1,\dots,K$ and $i=1,\dots,n$. We then calculate the integrated squared error for each time period $i$ as
\begin{equation*}
v_i = \int_t e_i^2(t)dt = \int_t \Big[\mathcal{X}_i(t) - \sum^K_{k=1}\widehat{\beta}_{i,k}\widehat{\phi}_k(t)\Big]^2dt.
\end{equation*}
This provides a measure of the estimation accuracy of the functional principal component decomposition for each period $i$. We then assign weights $w_i = 1$ if $v_i < s+ \lambda\sqrt{s}$ and 0 otherwise, where $s$ is the median of $\{v_1,\dots,v_n\}$ and $\lambda>0$ is a tuning parameter to control the degree of robustness. With these weights, we obtain updated estimates of $\widehat{\beta}_{i,k}$ and $\widehat{\phi}_k(t)$ using the Karhunen-Lo\`{e}ve decomposition. The value of $\lambda$ does not affect the optimal 50\% breakdown point, but it affects the efficiency of the algorithm. As noted by \cite{HU07}, $e_i(t)$ follows a normal distribution and $v_i$ follows a $\chi^2$ distribution with $\mathbb{E}(v_i) = \text{Var}(v_i)/2$. Using a normal approximation, $\text{Pr}(v_i<s+\lambda\sqrt{s})$ can be approximated by $\Phi(\lambda/\sqrt{2})$, where $\Phi$ denotes a cumulative normal distribution. When $\lambda=3$, the efficiency is $\Phi(3/\sqrt{2}) = 98.3\%$; this implies that 1.7\% of the total number of observations is classified as outliers. In practice, the optimal value of $\lambda$ may be estimated in a data-driven manner. In Section~\ref{sec:algorithm}, we present a computational algorithm for jointly selecting the optimal number of components and the optimal efficiency tuning parameter $\lambda$ in the robust FPCA.

Note that in \cite{HU07}, static functional principal components were extracted from the variance function alone. For modeling and forecasting functional time series that are moderate or strong dependent, the use of static FPCA may be inferior. Instead, our robust functional time series method estimates long-run covariance function, from which it extracts dynamic functional principal components. In Sections~\ref{sec:4} and~\ref{sec:5}, we show the superior forecast accuracy that dynamic FPCA entails, even when the data contain no outliers.
 
\subsection{A robust vector autoregressive model}\label{sec:robust_var}

The robust FPCA is more likely to result in robust functional principal components and their associated scores, but it may not completely remove all outliers because the proportion of outliers is unknown in practice. The un-detected outliers, manifested in the principal component scores, may still affect the estimation and forecasting performance of the chosen VAR model. The least squares estimates of $\bm{B}$ and $\bm{\Sigma}$ in~\eqref{eq:B_OLS} and~\eqref{eq:Var_OLS} could still be highly influenced by the presence of remaining outlying principal component scores affecting parameter estimates, model specification, and forecasts based on the VAR model. This motivates us to consider a robust VAR model to down-weigh the effect of outliers in the estimation and forecast of the principal component scores.

While the presence of outliers in multivariate time series analysis is not uncommon, it has not been studied extensively, apart from several noticeable exceptions. \cite{ACV08} replaced the multivariate least squares estimators in~\eqref{eq:B_OLS} and~\eqref{eq:Var_OLS} by an MLTS estimator; and \cite{MY13b} proposed a bounded MM-estimator for VAR models. Because of the availability of a computational algorithm (see \url{https://feb.kuleuven.be/public/u0017833/Programs/#mlts}), we use the robust procedure of \cite{ACV08} and compare its finite-sample performance with the OLS estimate of the VAR parameters in the presence and absence of residual outlying principal component scores.

Based on the idea of a minimum covariance determinant estimator, the MLTS estimator selects the subset of $\mathrm{h}$ observations having the property that the determinant of the covariance matrix of its residuals from least squares estimation, solely based on this subset, is minimal \citep{ACV08}. Let $\mathcal{H} = \{H\subset \{\omega+1,\dots,n\}|\# H=\mathrm{h}\}$ be the collection of all subsets of size $\mathrm{h}$. For any subset $H\in \mathcal{H}$, let $\widehat{\bm{B}}_{\text{OLS}}(H)$ be the classical least squares estimation based on the observations of the subset
\begin{equation*}
\widehat{\bm{B}}_{\text{OLS}}(H) = \left(\bm{X}_{H}^{\top}\bm{X}_{H}\right)^{-1}\bm{X}_H^{\top}\bm{Y}_H,
\end{equation*}
where $\bm{X}_H$ and $\bm{Y}_H$ are submatrices of $\bm{X}$ and $\bm{Y}$, having an index of $H$. The covariance matrix of error term computed from this subset is then
\begin{equation*}
\widehat{\bm{\Sigma}}_{\text{OLS}}(H) = \frac{\left[\bm{Y}_{H} -\bm{X}_H\widehat{\bm{B}}_{\text{OLS}}(H)\right]^{\top}\left[\bm{Y}_{H} -\bm{X}_H\widehat{\bm{B}}_{\text{OLS}}(H)\right]}{\mathrm{h}-(K+1)\omega-1}.
\end{equation*}
Consider the data pair $(\bm{x}_{\mathfrak{i}}, \bm{\beta}_{\mathfrak{i}}$), where $\bm{x}_{\mathfrak{i}} = \left(1,\bm{\beta}_{\mathfrak{i}-1}^{\top},\dots,\bm{\beta}_{\mathfrak{i}-\omega}^{\top}\right)^{\top}$ for $\mathfrak{i}=\omega+1,\dots,n$. The MLTS estimators of $\bm{B}$ and $H$ are defined as
\begin{align}
\widehat{\bm{B}}_{\text{OLS}}\big(\widehat{H}\big)&=\argmin_{\bm{B}, \bm{\Sigma}; |\bm{\Sigma}|=1}\sum^{\mathrm{h}}_{s=1}d^2_{s}(\bm{B},\bm{\Sigma}), \label{eq:B2_OLS} \\
\widehat{H}&=\argmin_{H\in \mathcal{H}}\text{det}\ \widehat{\bm{\Sigma}}_{\text{OLS}}(H) \notag
\end{align}
where $d_{1}(\bm{B},\bm{\Sigma})\leq \dots \leq d_{\mathrm{h}}(\bm{B},\bm{\Sigma})$ is the ordered sequence of the residual Mahalanobis distances
\begin{equation*}
d_i(\bm{B},\bm{\Sigma}) = \left[\left(\bm{\beta}_{\mathfrak{i}} - \bm{B}^{\top}\bm{x}_{\mathfrak{i}}\right)^{\top}\bm{\Sigma}^{-1}\left(\bm{\beta}_{\mathfrak{i}} - \bm{B}^{\top}\bm{x}_{\mathfrak{i}}\right)\right]^{\frac{1}{2}}
\end{equation*}
and the associated estimator of the covariance matrix is given by
\begin{equation}
\widehat{\bm{\Sigma}}_{\text{MLTS}}(\widehat{H}) = c_{\alpha}\times \widehat{\bm{\Sigma}}_{\text{OLS}}(\widehat{H}),\label{eq:Sigma_OLS}
\end{equation}
where $c_{\alpha}$ is a correction factor to obtain a consistent estimator of $\bm{\Sigma}$, and $\alpha$ represents the amount of trimming (i.e., $\alpha\approx 1-\mathrm{h}/n$ where $\mathrm{h}$ denotes the number of observations in the selected subset). Let $\mathfrak{q}$ be the dimension of the response variable. As demonstrated in \cite{CH99}, when error terms follow a multivariate normal distribution, $c_{\alpha} = (1-\alpha)/F_{\chi^2_{\mathfrak{q}+2}}(\mathfrak{q}_{\alpha})$, where $F_{\chi^2_{\mathfrak{q}+2}}$ denotes the cumulative distribution function of an $\chi^2$ distribution with $\mathfrak{q}+2$ degrees of freedom, and $\mathfrak{q}_{\alpha} = \chi^2_{\mathfrak{q},1-\alpha}$ represents the upper $\alpha$-quantile.

The efficiency of the MLTS estimator can be further improved by one-step re-weighing. Let $\widehat{\bm{B}}_{\text{MLTS}}$ and $\widehat{\bm{\Sigma}}_{\text{MLTS}}$ denote the initial MLTS estimates obtained in~\eqref{eq:B2_OLS} and~\eqref{eq:Sigma_OLS}. Then, the one-step re-weighed multivariate least trimmed squares (RMLTS) estimates are given as
\begin{align*}
\widehat{\bm{B}}_{\text{RMLTS}} &= \widehat{\bm{B}}_{\text{OLS}}(J) \\
\widehat{\bm{\Sigma}}_{\text{RMLTS}} &= c_{\delta}\times\widehat{\bm{\Sigma}}_{\text{OLS}}(J),
\end{align*}
where 
\begin{equation}
J=\left\{j\in \{1,\dots,n\}\Big|d_j^2(\widehat{\bm{B}}_{\text{MLTS}}, \widehat{\bm{\Sigma}}_{\text{MLTS}})\leq \mathfrak{q}_{\delta}\right\}, \qquad \mathfrak{q}_{\delta} = \chi^2_{q,1-\delta}, \label{eq:J(K)}
\end{equation}
where $\mathfrak{q}_{\delta} = \chi^2_{\mathfrak{q},1-\delta}$ represents the upper $\delta$-quantile of a $\chi^2$ distribution. Here $\delta$ denotes the trimming proportion, and
\begin{equation*}
c_{\delta} = \frac{1-\delta}{\int_{\|u\|^2\leq q_{\delta}}u_1^2dF_0(u)}
\end{equation*}
is a consistency factor to achieve Fisher consistency at the modal distribution \citep[see][]{ACV08}. In the case of multivariate normal errors, we have
\begin{equation*}
c_{\delta} = \frac{1-\delta}{F_{\chi^2_{\mathfrak{q}+2}}(\mathfrak{q}_{\delta})}.
\end{equation*}

If the squared Mahalanobis distance is larger than the critical value, the observations are flagged as outliers. \cite{ACV08} considered $\delta=0.01$ and take the trimming proportion of the initial MLTS estimator $\alpha=25\%$. In the simulation studies and data applications, we introduce an optimization algorithm, described in Section~\ref{sec:algorithm}, for selecting the optimal values of $\alpha$ and $\delta$.

\subsection{Robust order selection in vector autoregressive models}\label{sec:robust_order}

Determination of the model order is an important step in any VAR modeling, given that the number of parameters grows very rapidly with the lag length. To select the optimal order, information criteria strike a compromise between lag length and number of parameters by minimizing a linear combination of the residual sum of squares and the number of parameters. The most common information criterion includes the AIC, BIC, or Hannan-Quinn criterion. We consider the BIC criterion. These information criteria require the computation of log likelihood and a penalty term for penalizing model complexity. Under the form of error distribution in~\eqref{eq:error_den}, we obtain
\begin{equation}
l_{\omega}= \sum^n_{\mathfrak{i}=\omega+1}g\left(\bm{u}_{\mathfrak{i}}^{\top}\bm{\Sigma}^{-1}\bm{u}_{\mathfrak{i}}\right) - \frac{n-\omega}{2} \times \ln \text{det} \bm{\Sigma}, \label{eq:order_selection}
\end{equation}
where $\bm{u}_{\mathfrak{i}}$ is the mean of the $\mathfrak{i}^{\text{th}}$ error term, and $\bm{\Sigma}$ is a positive definite matrix of the error term. When $g(\cdot)$ follows a multivariate normal distribution,~\eqref{eq:order_selection} can be expressed as
\begin{equation}
l_{\omega} = -\frac{(n-\omega)K}{2}\times  \ln (2\pi) - \frac{1}{2}\sum^n_{\mathfrak{i}=\omega+1}\bm{u}_{\mathfrak{i}}^{\top}\bm{\Sigma}^{-1}\bm{u}_{\mathfrak{i}} - \frac{n-\omega}{2}\times \ln \text{det} \bm{\Sigma}. \label{eq:likelihood}
\end{equation}
Although $\bm{\Sigma}$ is unknown, it can be estimated by the OLS
\begin{equation}
\widehat{\bm{\Sigma}}_{\text{OLS}} = \frac{\sum^n_{\mathfrak{i}=\omega+1}\widehat{\bm{u}}_{\mathfrak{i}}\widehat{\bm{u}}_{\mathfrak{i}}^{\top}}{n-(K+1)\omega-1}. \label{eq:sigma_OLS}
\end{equation}
Via the RMLTS estimator, the variance $\bm{\Sigma}$ can also be estimated by
\begin{equation}
\widehat{\bm{\Sigma}}_{\text{RMLTS}} = \frac{c_{\delta}\sum_{\mathfrak{i}\in J}\widehat{\bm{u}}_{\mathfrak{i}}\widehat{\bm{u}}_{\mathfrak{i}}^{\top}}{\mathrm{m}-(K+1)\omega-1}, \label{eq:sigma_RMLTS}
\end{equation}
where $J$ is defined in~\eqref{eq:J(K)}, and $\mathrm{m}$ denotes the number of elements in $J$. By plugging in~\eqref{eq:sigma_OLS} and~\eqref{eq:sigma_RMLTS} into~\eqref{eq:likelihood} and adding the penalty term, we obtain
\begin{align*}
\text{Criterion} =&\ -2\times \frac{l_{\omega}}{n-\omega} + \ln(n-\omega)\times \frac{Kq}{n-\omega} \\
=&\ \frac{(n-\omega)\times \ln \text{det}\Sigma + (n - \omega)K\times \ln (2\pi) + \sum^n_{\mathfrak{i}=\omega+1}\bm{u}_{\mathfrak{i}}^{\top}\bm{\Sigma}^{-1}\bm{u}_{\mathfrak{i}}}{n - \omega} + \ln(n-\omega) \times \frac{Kq}{n-\omega} \\
=& \ \ln \text{det} \bm{\Sigma} + K \ln (2\pi) + \frac{1}{n-\omega}\sum^n_{\mathfrak{i}=\omega+1}\bm{u}_{\mathfrak{i}}^{\top}\bm{\Sigma}^{-1}\bm{u}_{\mathfrak{i}} + \ln(n-\omega) \times \frac{Kq}{n-\omega}, 
\end{align*}
where $Kq=K(K\omega+1)$ is the number of unknown parameters, which penalizes model complexity. The optimal VAR model is the one with the minimum BIC.

\subsection{Computational algorithm}\label{sec:algorithm}

We summarize the steps involved in obtaining robust functional principal components and their associated principal component scores, and the steps involved in obtaining forecast principal component scores and in turn forecast curves. The steps are listed as follows:
\vspace{.1in}

\begin{asparaenum}
\item[1)] We apply the robust FPCA of \cite{HU07} to obtain robust functional principal components and their associated scores. These scores can be modeled and forecasted via a VAR model, where the parameters are estimated via the OLS method. Based on a validation data set, we determine the optimal number of components $K$ described in Section~\ref{sec:21}, and the efficiency of tuning parameter $\lambda$ described in Section~\ref{sec:3.1} by minimizing a forecast error measure, such as mean squared forecast error (MSFE), averaged over all data in the validation data set. It is important to optimize an integer-valued parameter $K$ and a positive real-valued parameter $\lambda$ jointly. Given $K$ is an integer, we consider $K$ between 1 and 50. Furthermore, we divide $K$ by 50, so that the transformed parameter is a proportion bounded between 0.02 and 1. By taking the logit transformation, we obtain a new parameter that lies within the real-valued parameter space. Then, we use an optimization algorithm, such as \cite{NM65}, to obtain the optimal value for a parameter vector. For $K$, we then take inverse logit transformation to obtain its optimal integer.
\vspace{.1in}
\item[2)] Having removed the outlying functions identified by robust FPCA, we compute an estimate of the long-run covariance function. With the estimated dynamic principal component scores, we can also estimate the parameters in a VAR model by the proposed MLTS and RMLTS estimators to model and forecast these scores. Apart from jointly selecting $K$ and $\lambda$ used in the robust FPCA, we have one or two additional parameters depending on if the MLTS or RMLTS estimator described in Section~\ref{sec:robust_var} is implemented. We consider to select the optimal truncation parameter $\alpha$ in the MLTS estimator and optimal re-weighing parameter $\delta$ in the RMLTS estimator jointly, along with $K$ and $\lambda$ by minimizing the averaged MSFE in the validation data set. 
\vspace{.1in}
\item[3)] With the one-step-ahead forecast of principal component scores $\widehat{\bm{\beta}}_{n+h}$, we compute the one-step-ahead forecasts of functional curves $\mathcal{X}_{n+h}(t)$ by multiplying the forecast principal component scores $\widehat{\bm{\beta}}_{n+h}$ with the estimated functional principal components $\widehat{\bm{\phi}}(t)$ from the four methods.
\end{asparaenum}

\section{Simulation study}\label{sec:4}

\begin{figure}[!htbp]
\centering
\includegraphics[width=11cm]{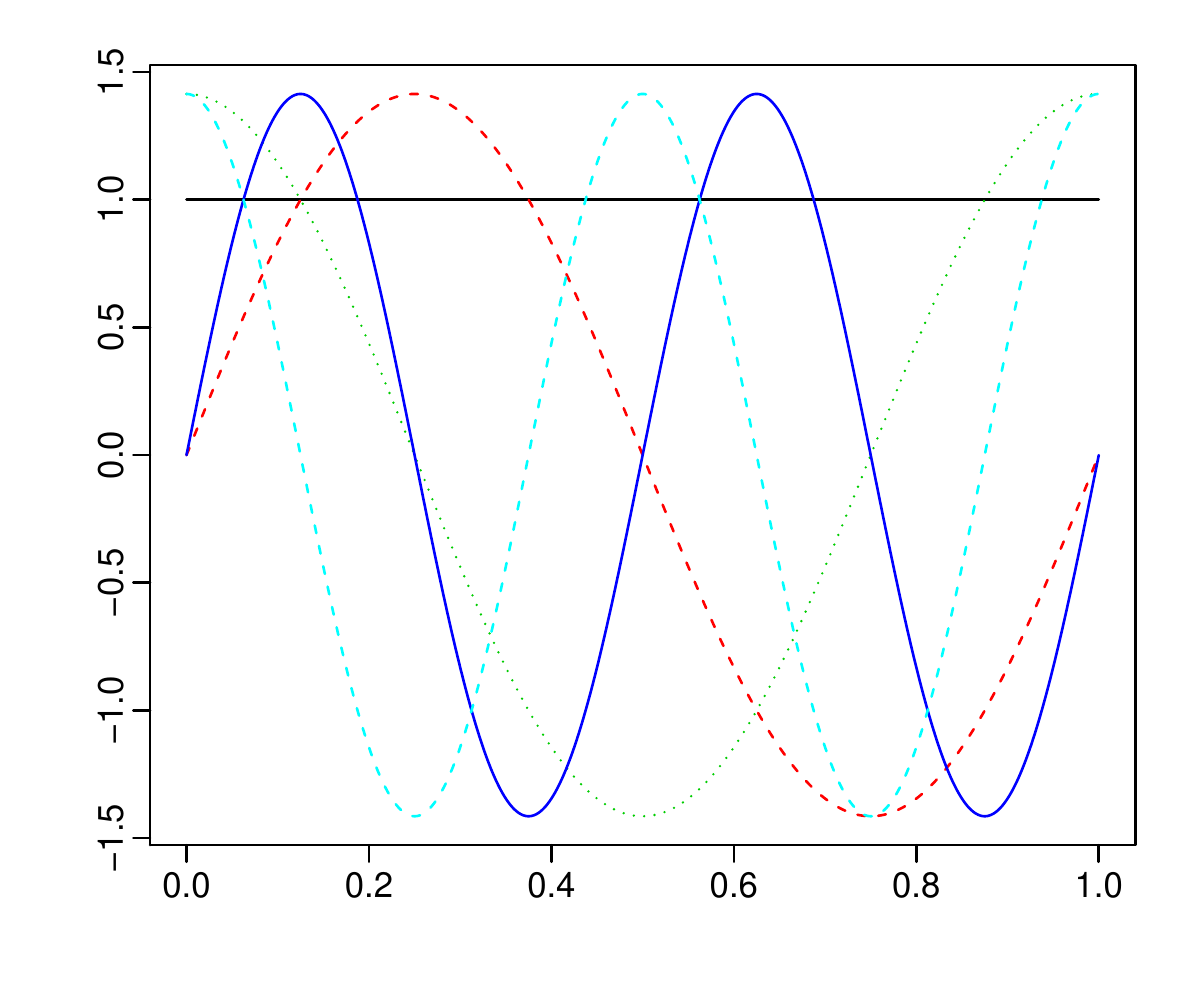}
\caption{The first five Fourier basis functions on unit interval.}\label{fig:Fourier}
\end{figure}

To analyze the finite-sample performance of the new prediction method, a comparative simulation study was conducted. We consider a functional autoregressive of order 1 (FAR(1)) process previously studied in \cite{ANH15}, where the functional curves are simulated from a kernel operator. The simulation setup consists of $D$ Fourier basis functions $\left\{\nu_1,\dots,\nu_{D}\right\}$ on the unit interval $[0,1]$, which jointly determine the (finite-dimensional) space $H = \text{sp}\{\nu_1,\dots,\nu_{D}\}$. In Figure~\ref{fig:Fourier}, we present the first five Fourier basis functions as an illustration.

Innovations were defined by setting
\begin{equation*}
\epsilon_i(t) = \sum^{D}_{l=1}A_{i,l}\nu_l(t),
\end{equation*}
where $(A_{i,1},\dots, A_{i,D})^{\top}$ were iid random vectors with mean zero and standard deviation $\sigma_l$. The data generating process we consider is:
\begin{equation*}
\text{FAR}(1): \quad \mathcal{X}_i = \Psi(\mathcal{X}_{i-1}) + \epsilon_i.
\end{equation*}

To generate the functional autoregressive time series, we set the starting value 
\begin{equation*}
\mathcal{X}_{-9} = \sum^{D}_{l=1}N_l\nu_l
\end{equation*}
with a normal random vector $(N_1,\dots,N_{D})^{\top}\sim N(\bm{0},\bm{I}_{D})$. The first ten elements $(\mathcal{X}_{-9},\dots,\mathcal{X}_0)$ were used for a burn-in.

It is noteworthy that an arbitrary element in Hilbert space $\mathscr{H}$ has the following representation:
\begin{equation*}
\mathcal{X}(t) = \sum^{D}_{l=1}c_l\nu_l(t),
\end{equation*}
with coefficients $\bm{c} = (c_1,\dots,c_{D})^{\top}$. If $\Psi: \mathscr{H}\rightarrow \mathscr{H}$ is a linear operator, then
\begin{align*}
\Psi(x) &= \sum^{D}_{\ell=1}c_{\ell}\Psi(v_{\ell}) \\
&= \sum^{D}_{\ell=1}\sum^{D}_{\ell^{'}=1}c_{\ell}\langle\Psi(v_{\ell}), v_{\ell^{'}}\rangle v_{\ell^{'}} \\
&= (\bm{\Psi}\bm{c})^{\top}\bm{v},
\end{align*}
where $\bm{\Psi}$ is the matrix whose $\ell^{'}$-th row and $\ell$-th column is $\langle \bm{\Psi}(v_{\ell}), v_{\ell^{'}} \rangle$ and $\bm{v}=(v_1,\dots,v_{10})^{\top}$ is the vector of basis functions. The linear operator can be represented by a $D\times D$ matrix that operates on the coefficients in the basis function representation of the curves. 

For the purpose of demonstration, let $D=3$ and $\sigma_1 = \sigma_2 = \sigma_3 = 1$. The autocovariance operator $\bm{\Psi}$ with corresponding matrix
\vspace{.05in}
\begin{equation*}
\bm{\Psi} = \begin{pmatrix*}[r]
-0.05 & -0.23 & 0.76 \\
0.80 & -0.05 & 0.04 \\
0.04 & 0.76 & 0.23 
\end{pmatrix*} 
\end{equation*}
was tested. At lags greater than 1, there is a considerable dependence in the cross-correlations, thus it is advantageous to use a VAR model instead of a univariate time series model.

We randomly simulate 200 functional curves with different amounts of contamination and keep the last 80 curves as a testing sample. To select the optimal tuning parameters, we again minimize the averaged MSFE over a validation data set consisting of observations 61 to 120. Using the first 120 simulated curves as the initial training sample, we estimate parameters in the VAR model based on the (robust) BIC. Then, we produce a one-step-ahead point forecast, evaluate and compare its point forecast accuracy with the corresponding data in the testing sample. 

\begin{figure}[!htbp]
\centering
\includegraphics[width=18.5cm]{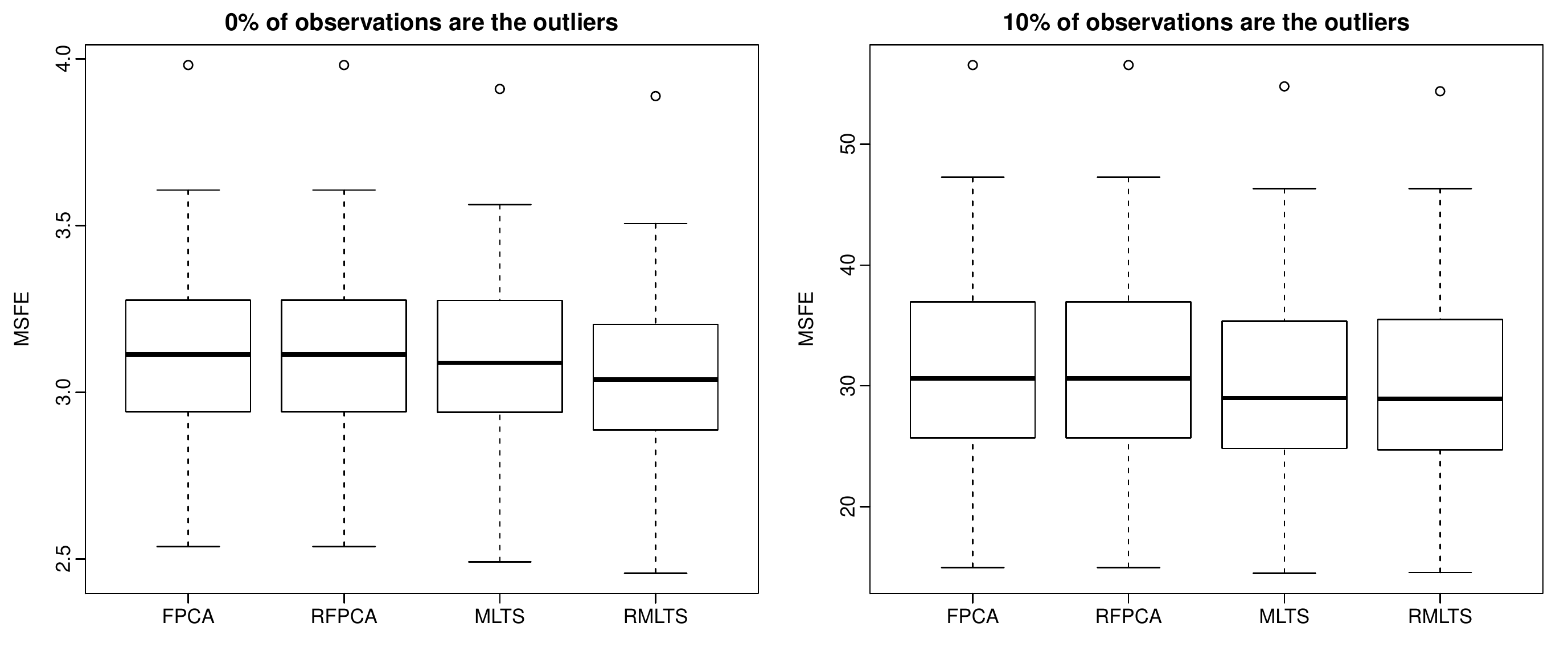}
\caption{Comparison of point forecast accuracy between the standard FPCA and robust FPCA (abbreviated by RFPCA), where the testing sample may contain outliers. For the principal component scores obtained from the robust FPCA, the point forecast accuracy of the optimal VAR model is further compared among the OLS, MLTS and RMLTS estimators.}\label{fig:2}
\end{figure}

We repeat our simulation setup for 100 replications with different pseudorandom seeds, and then use the summary statistics of MSFEs to evaluate and compare one-step-ahead forecast accuracy. From Figure~\ref{fig:2}, the performance of the RMLTS estimator gives the best estimation accuracy, when there is no outlier in the training sample or when there is 10\% of observations that are outliers. This result shows the advantage of dynamic functional principal components as basis functions for forecasting, coupled with the robust VAR forecasting method.

\begin{figure}[!htbp]
\centering
\includegraphics[width=18.5cm]{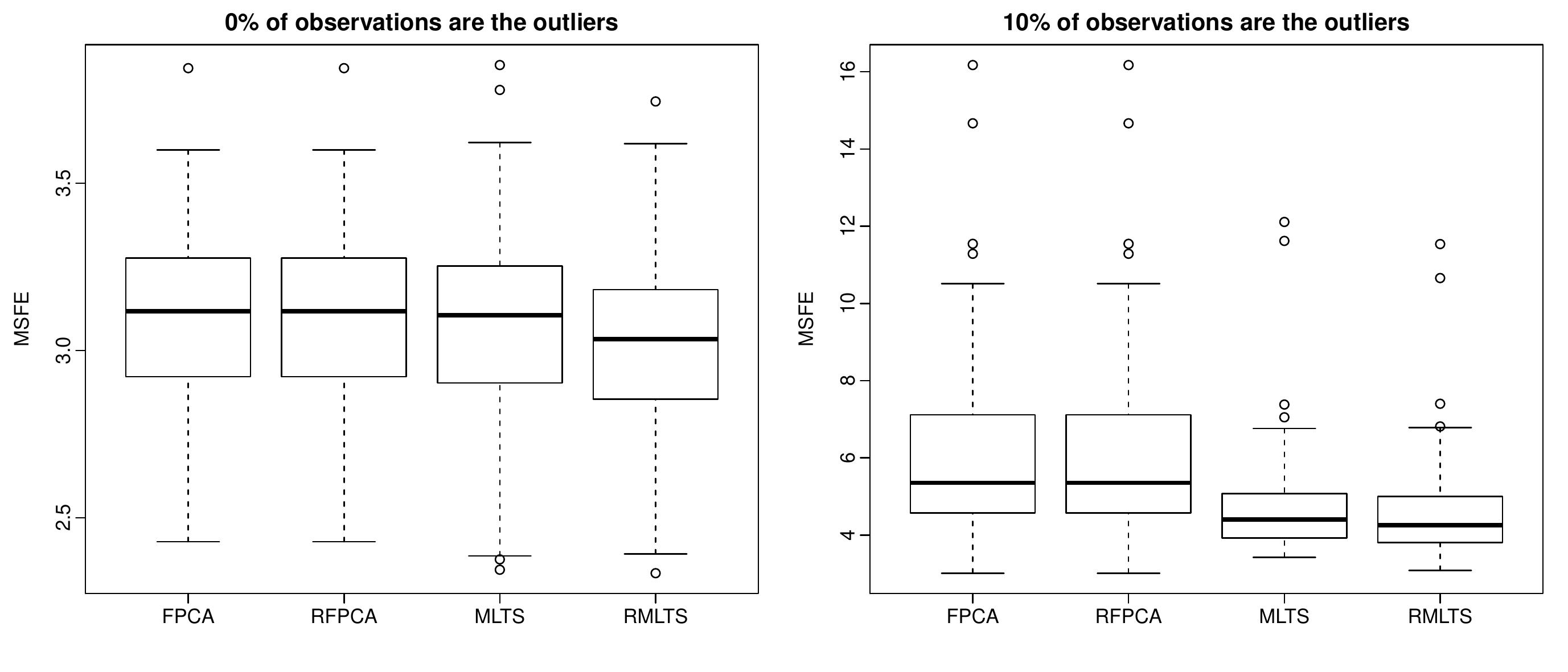}
\caption{Comparison of point forecast accuracy between the standard FPCA and robust FPCA (abbreviated by RFPCA), where the testing sample is free from outliers.}\label{fig:3}
\end{figure}

To study the increase in MSFE that due to contamination in the testing sample, forecast accuracy should also be evaluated against ``true" observations that are cleaned from contamination. We repeat our simulation setup for 100 replications with different pseudo random seeds, where the testing sample is free from contamination. We use the summary statistics of MSFEs to evaluate and compare one-step-ahead point forecast accuracy. From Figure~\ref{fig:3}, the performance of the RMLTS estimator gives the best estimation accuracy. 

In terms of computational speed, the fastest and easiest to use method is the standard FPCA. Based on one simulated sample with $n=200$ curves, we implemented the four methods to produce one-step-ahead point forecasts and calculate their point forecast errors for one testing sample. The computational speeds are listed below:
\begin{table}[!htbp]
\tabcolsep 0.78in
\centering
\caption{Computational speed measured in second on a MacBook Pro with 2GHz Intel Core i7 and 16GB 1600 MHz DDR3 memory.}
\begin{tabular}{@{}llll@{}}
\toprule
FPCA & Robust FPCA & \multicolumn{2}{c}{Robust dynamic FPCA} \\
	  & OLS & MLTS & RMLTS \\
	  \midrule
72.36 & 74.34 & 76.60 & 77.39 \\
\bottomrule
\end{tabular}
\end{table}

\section{Ozone pollution data}\label{sec:5}

Ground-level ozone is an air pollutant known to cause serious health problems. High levels of O$_3$ can produce harmful effects on human health and the environment. The World Health Organization, in the 2005 global update of its quality guidelines (World Health Organization, 2006), reduced the guideline levels of ozone from 120 ug$m^{-3}$ (8-hour daily average) to 100 ug$m^{-3}$ for a daily maximum 8-hour mean. Ozone levels above this threshold can harm human health. Thus, modeling ground-level ozone formation has been an active topic of air quality studies for many years. Following an early study by \cite{Gervini12}, we consider a data set, obtained from the California Environmental Protection Agency (available at \url{https://www.arb.ca.gov/aqd/aqdcd/aqdcddld.htm}), on hourly concentrations of O$_3$ at different locations in California for the years from 1980 to 2009. Here, we will focus on the trajectories of ozone (O$_3$) in the city of Sacramento (site 3011 in the database). There are nine days with some missing observations, so we impute the missing data via linear interpolation. Following \cite{Gervini12}, we also applied a square-root transformation to stabilize extreme observations. 

\begin{figure}[!htbp]
\centering
\subfloat[A univariate time series display]
{\includegraphics[width=8.7cm]{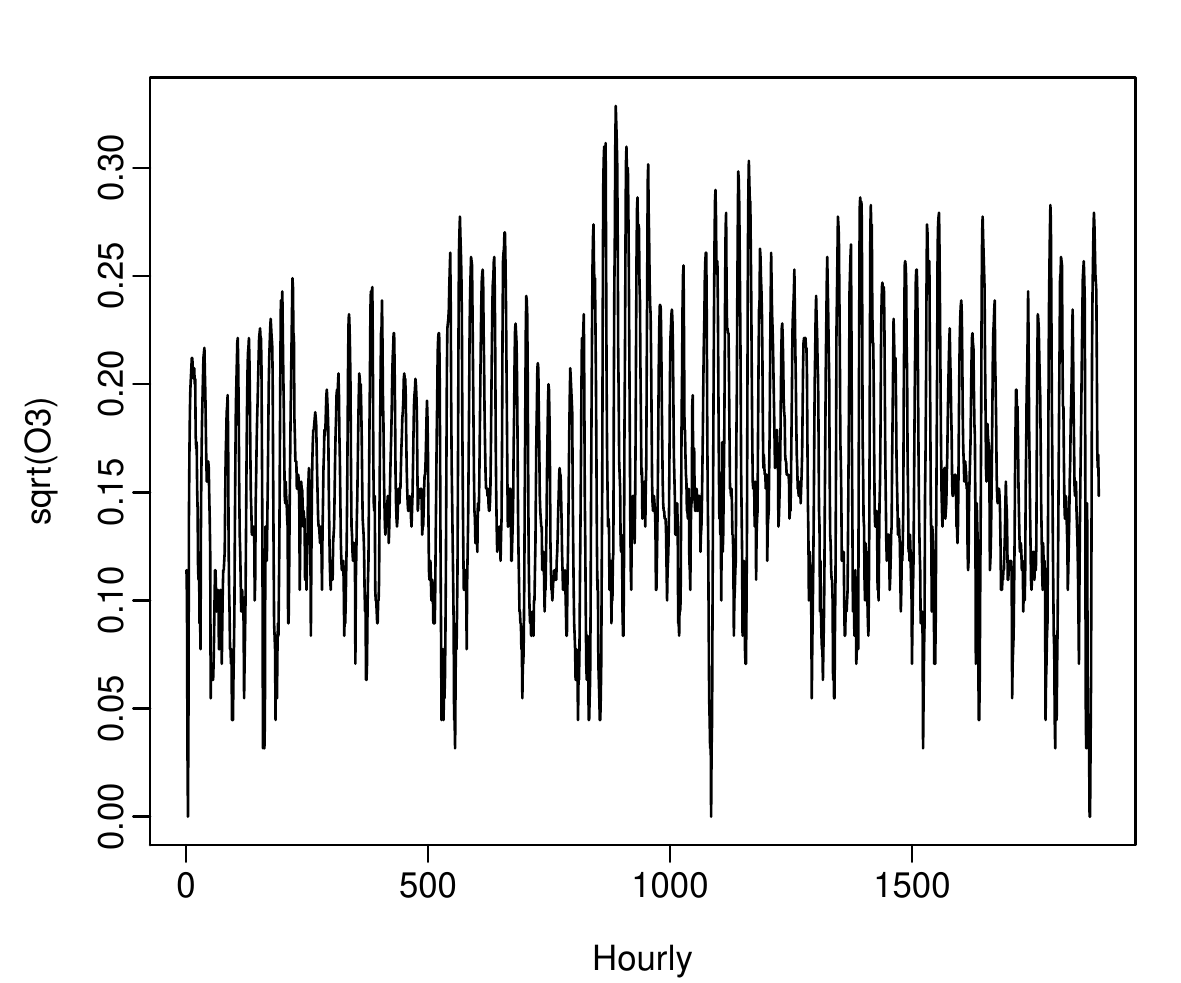}\label{fig:1a}}
\qquad
\subfloat[A functional time series display]
{\includegraphics[width=8.7cm]{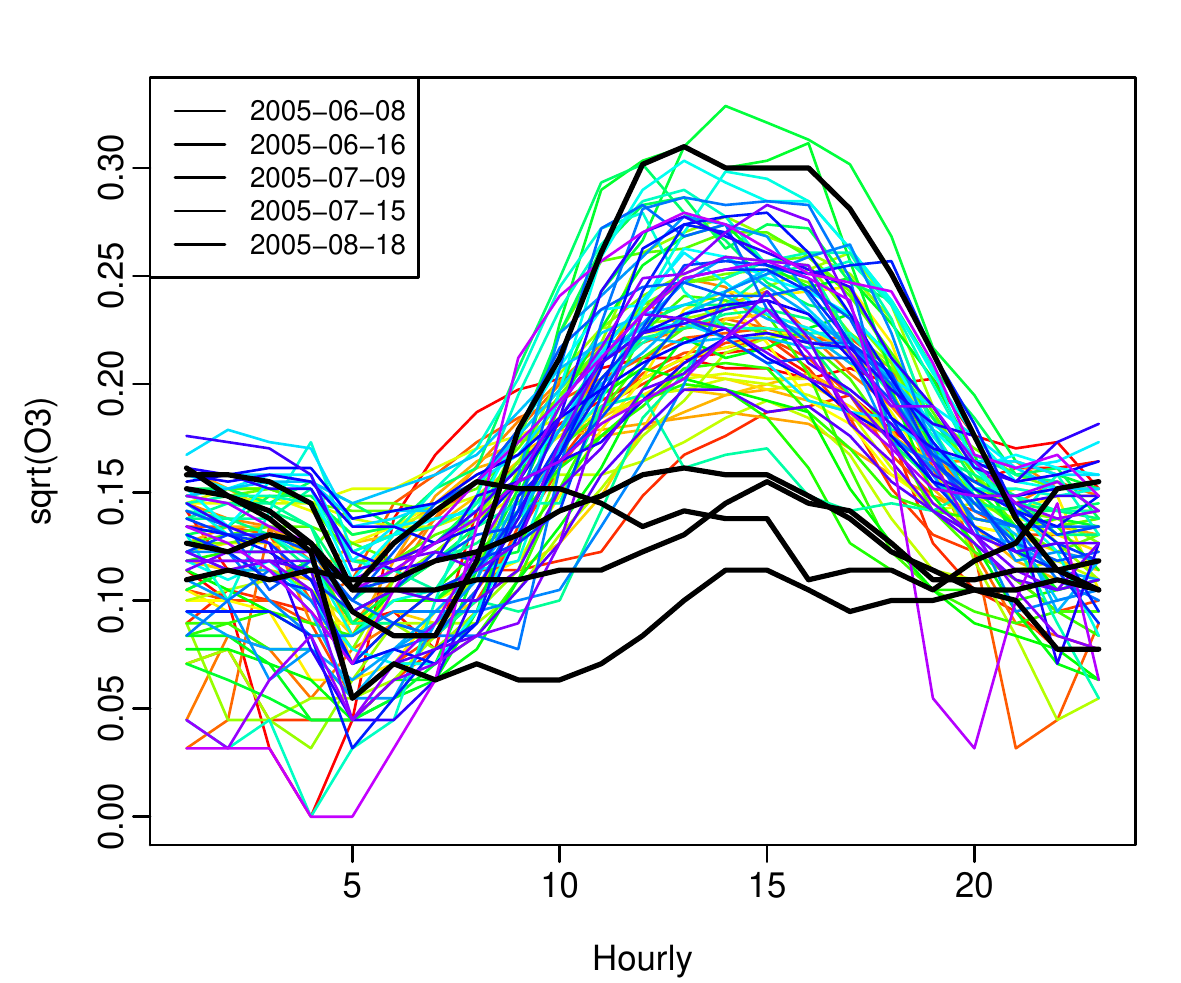}\label{fig:1b}}
\caption{Graphical displays of hourly ozone concentration from June 6, 2005 to August 26, 2005. Five detected outliers are highlighted by the thick black lines.}
\end{figure}

A univariate time series display of hourly ozone concentration is given in Figure~\ref{fig:1a}, with the same data shown in Figure~\ref{fig:1b} as a time series of functions. From Figure~\ref{fig:1b}, there are some years showing extreme measurements of ozone concentration, which are suspected to be outliers. Given that the presence of outliers can seriously hinder the performance of modeling and forecasting, outlier detection for functional data has recently received increasing attention in the literature \citep[e.g.][]{HS10}. By using a functional highest density region (HDR) boxplot, we detected five outliers corresponding to dates June 8, June 16, July 9, July 15 and August 18 in 2005 (highlighted by the thick black lines in Figure~\ref{fig:1b}). These outliers may due to high temperatures, low winds, and an inversion layer. However, it is difficult to verify the correctness of outlier-detection accuracy in practice. Instead of removing outliers, we evaluate and compare the point forecast accuracy between the standard and robust functional time series forecasting methods. 

\begin{figure}[!htbp]
\centering
\includegraphics[width=10.5cm]{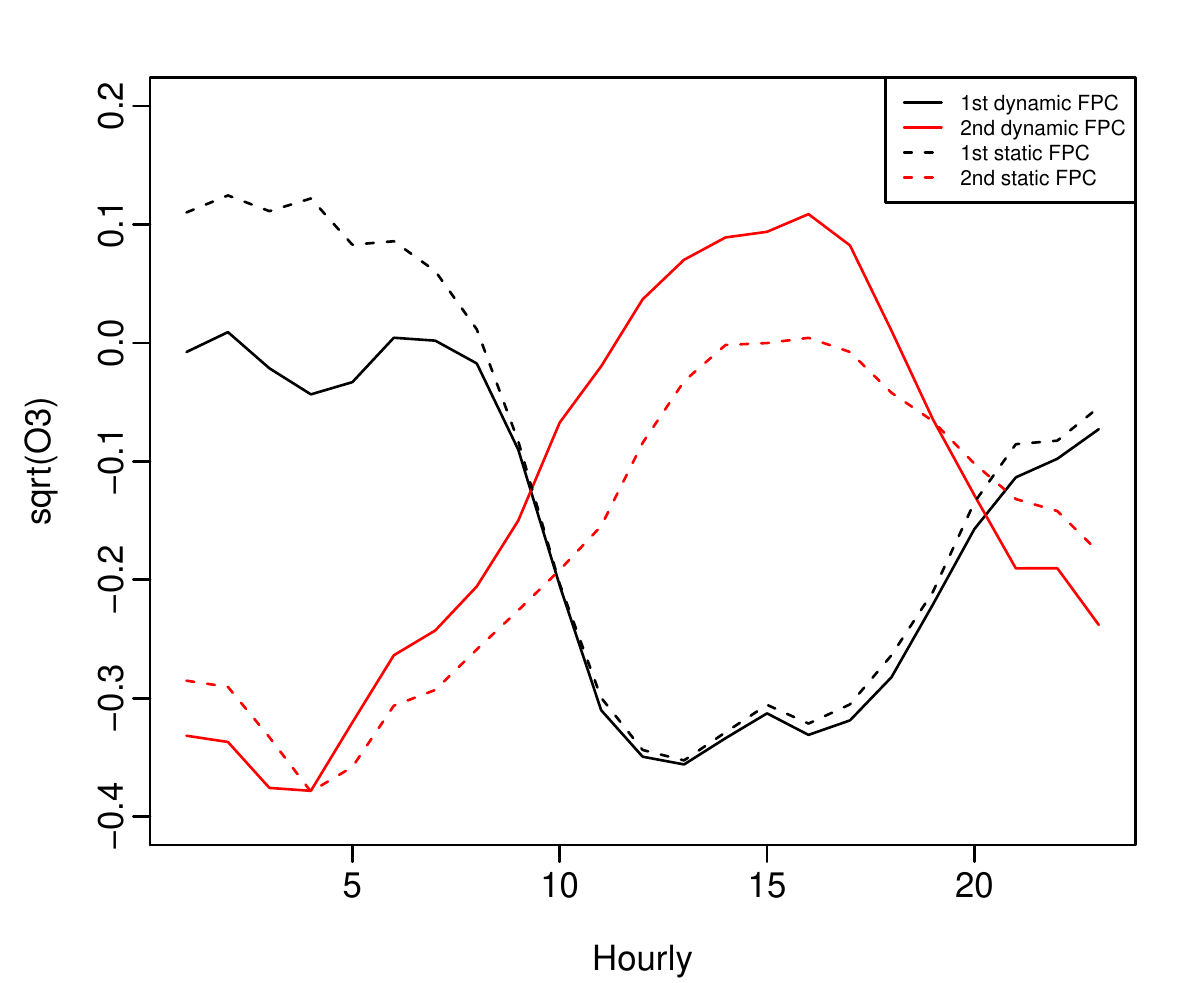}
\caption{A comparison between the first two dynamic and static functional principal components.}\label{fig:5}
\end{figure}

In Figure~\ref{fig:5}, we present a visual comparison between the first two dynamic and static functional principal components. The dynamic components were extracted from the estimated long-run covariance function, while the static components were extracted from the estimated variance function. The former one can capture additional information when functional time series exhibits a moderate or strong dependence. For this dataset, the first two dynamic functional principal components explain 59\% and 26\% of the total variation, while the first two static functional principal components explain 48\% and 24\% of the total variation. 

It is noteworthy that the robust forecasting method is designed to down-weigh the influence of outliers among observations to produce a forecast that does not differ much from a ``normal" observation. If an actual observation that we attempt to predict happens to be an outlier, then it is plausible that robust forecasting method may perform worse. However, on average, the robust forecasting methods perform better than the non-robust forecasting methods, when observations in both the fitting and forecasting periods contain outliers.

\subsection{Point forecast evaluation setup}\label{sec:5.1}

We split our data into a training sample (including data from day 1 to $(n-33)$) and a testing sample (including data from day $(n-32)$ to $n$), where $n=82$ represents the total number of curves. We implement an expanding window approach, which is commonly used to assess model and parameter stabilities over time. With the initial training sample, we produce one-day-ahead forecasts and determine the point forecast errors by comparing the forecasts with the holdout samples. As the training sample increases by one year, we again produce one-day-ahead forecasts and calculate the point forecast errors. This process continues until the training sample covers all available data. We compare these forecasts with the holdout samples to determine the out-of-sample point forecast accuracy.

We further split the training sample into a training set (including data from day 1 to $(n-57)$) and a validation set (including data from day $(n-56)$ to $(n-33)$). Using an expanding window approach, we determine the optimal tuning parameters based on the smallest MSFE averaged over the data in the validation set.

\subsection{Point forecast accuracy results}\label{sec:4.2}

Using the point forecast evaluation setup in Section~\ref{sec:5.1}, we examine the one-step-ahead MSFE obtained from the functional principal component regression models. For the data in the forecasting period, the optimal order of the VAR model is VAR(1) selected by both standard and robust BIC described in Section~\ref{sec:robust_order}. We compare the point forecast errors between the standard and robust functional principal component decompositions. In the latter approach, we compare three methods where the principal component scores can be modeled and forecasted by a VAR model using the OLS, MLTS and RMLTS estimators. 

\begin{figure}[!htbp]
\centering
\includegraphics[width=11cm]{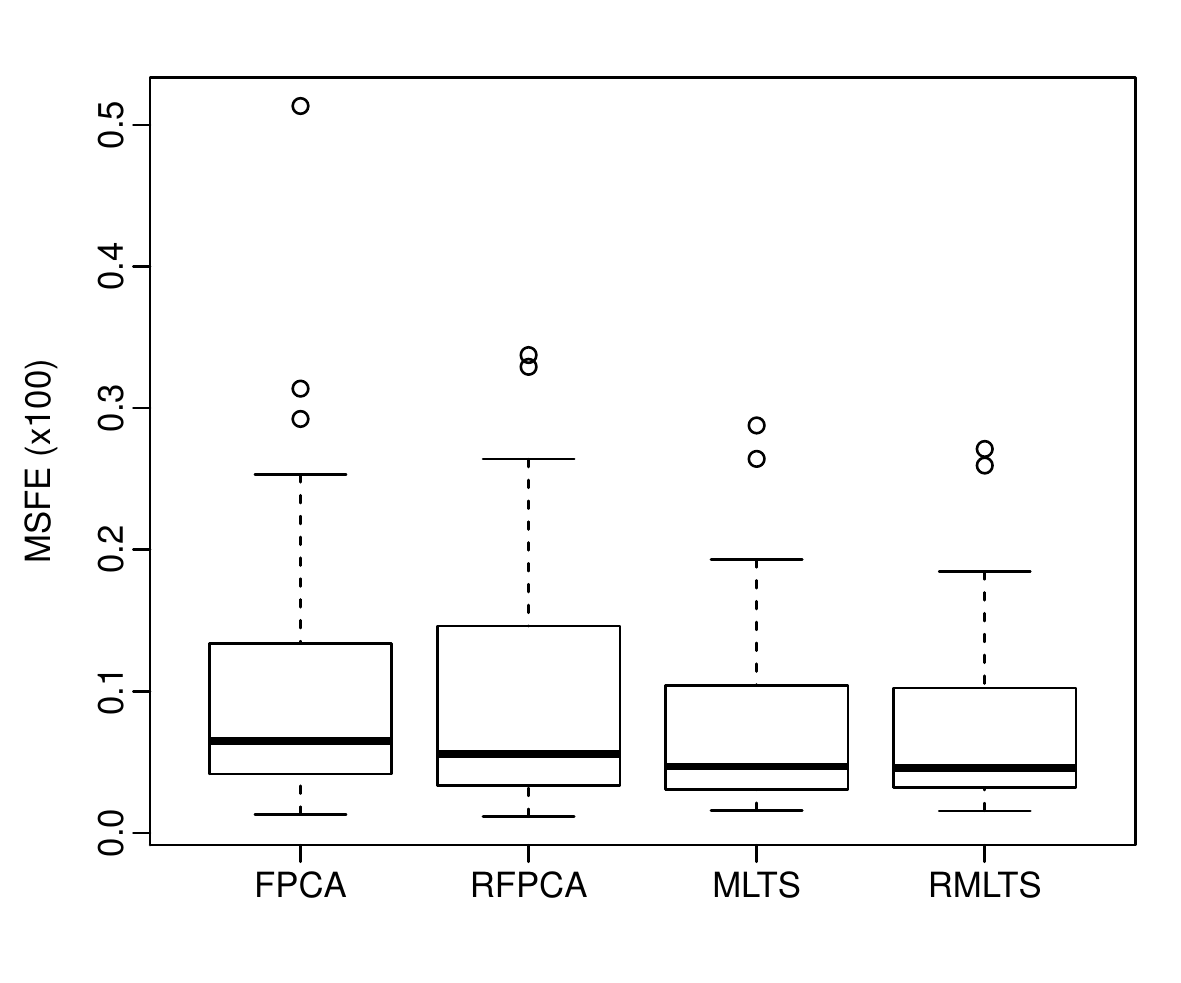}
\caption{Comparison of point forecast accuracy between the standard FPCA and robust FPCA (abbreviated by RFPCA).}\label{fig:1point}
\end{figure}

In Figure~\ref{fig:1point}, we show the point forecast accuracy, as measured by the MSFE for all data in the forecasting period. The robust FPCA performs similarly to the standard FPCA. This similarity stems from the fact that we only allow 1.7\% of observations to be classified as outliers ($\lambda = 3$ by default; see Section~\ref{sec:3.1}). There may still be outliers remaining that could affect the forecast of principal component scores. This motivates us to consider the robust estimators for the optimal VAR model, which give the best point forecast accuracy among the methods we considered. Between the MLTS and RMLTS estimators, the RMLTS estimator performs slightly better than the MLTS estimator as shown in Table~\ref{tab:2}. 

\begin{table}[!htbp]
\tabcolsep 0.35in
\centering
\caption{Summary statistics of the MSFEs ($\times 100$) among the four estimators.}\label{tab:2}
\begin{tabular}{@{}lrrrr@{}}
\toprule
 &FPCA & Robust FPCA & \multicolumn{2}{c}{Robust dynamic FPCA} \\
 & & & MLTS & RMLTS \\ 
\midrule
Min. &  0.0131 & 0.0115 & 0.0158 & 0.0156 \\ 
  1st Qu.  & 0.0418 & 0.0335 & 0.0307 & 0.0321 \\ 
  Median  & 0.0651 & 0.0559 & 0.0470 & 0.0457 \\ 
  Mean & 0.1063 & 0.0961 & 0.0782 & 0.0763 \\ 
  3rd Qu. & 0.1337 & 0.1461 & 0.1042 & 0.1024 \\ 
  Max. & 0.5134 & 0.3375 & 0.2878 & 0.2712 \\ 
  sd & 0.1085 & 0.0916 & 0.0703 & 0.0670  \\
\bottomrule
\end{tabular}
\end{table}

\subsection{Statistical significance test} \label{sec:5.3}

To examine statistical significance based on out-of-sample point forecast accuracy, we implement the model confidence set (MCS) procedure proposed by \cite{HLN11}. It consists of a sequence of tests which permits to construct a set of ``superior" models, where the null hypothesis of equal predictive ability (EPA) is not rejected at a certain confidence level. The EPA test statistic can be evaluated for any arbitrary loss function, such as absolute loss function or squared loss function considered here.

Let $M$ be some subset of $M^0$ and let $m=4$ be the number of candidate models in $M$, and let $d_{\rho \xi,i}$ denotes the loss differential between two models $\rho$ and $\xi$, that is
\begin{equation*}
d_{\rho \xi,i} = l_{\rho, i} - l_{\xi, i}, \qquad \rho, \xi=1,\dots,m, \quad i=1,\dots,n,
\end{equation*}
and calculate
\begin{equation*}
d_{\rho\cdot, i} = \frac{1}{m}\sum_{\xi\in M}d_{\rho \xi, i}, \qquad \rho = 1,\dots,m
\end{equation*}
as the loss of model $\rho$ relative to any other model $\xi$ at time point $i$. The EPA hypothesis for a set of $M$ candidate models can be formulated in two ways:
\begin{align}
\text{H}_{0,M}: c_{\rho \xi}&=0, \quad \text{or}\quad c_{\rho.}=0 \qquad \text{for all}\quad \rho, \xi = 1,2,\dots,m\label{eq:hypo_1}\\
\text{H}_{A,M}: c_{\rho \xi}&\neq 0, \quad \text{or}\quad c_{\rho.}\neq 0 \qquad \text{for some}\quad \rho, \xi = 1,2,\dots,m.\label{eq:hypo_2}
\end{align}
where $c_{\rho \xi} = \mathbb{E}(d_{\rho \xi})$ and $c_{\rho.} = \mathbb{E}(d_{\rho.})$ are assumed to be finite and not time dependent. Based on $c_{\rho\xi}$ or $c_{\rho.}$, we construct two hypothesis tests as follows:
\begin{equation}
t_{\rho \xi} = \frac{\overline{d}_{\rho \xi}}{\sqrt{\widehat{\text{Var}}(\overline{d}_{\rho \xi})}}, \qquad t_{\rho.} = \frac{\overline{d}_{\rho.}}{\sqrt{\widehat{\text{Var}}(\overline{d}_{\rho.})}}, \label{eq:t_ij_t_i}
\end{equation}
where $\overline{d}_{\rho.} = \frac{1}{m}\sum_{\xi\in M}\overline{d}_{\rho \xi}$ is the sample loss of $\rho^{\text{th}}$ model compared to the averaged loss across models, and $\overline{d}_{\rho \xi} =\frac{1}{n}\sum^n_{i=1}d_{\rho \xi,i}$ measures the relative sample loss between the $\rho^{\text{th}}$ and $\xi^{\text{th}}$ models. Note that $\widehat{\text{Var}}(\overline{d}_{\rho.})$ and $\widehat{\text{Var}}(\overline{d}_{\rho \xi})$ are the bootstrapped estimates of $\text{Var}(\overline{d}_{\rho.})$ and $\text{Var}(\overline{d}_{\rho \xi})$, respectively. \cite{BC14} perform a block bootstrap procedure with 5,000 bootstrap samples by default, where the block length $p$ is given by the maximum number of significant parameters obtained by fitting an AR($p$) process on all the $d_{\rho \xi}$ term. For both hypotheses in~\eqref{eq:hypo_1} and~\eqref{eq:hypo_2}, there exist two test statistics:
\begin{equation*}
T_{\text{R}, \text{M}} = \max_{\rho,\xi\in M}|t_{\rho \xi}|, \qquad T_{\max, \text{M}} = \max_{\rho\in M} t_{\rho.},
\end{equation*} 
where $t_{\rho \xi}$ and $t_{\rho.}$ are defined in~\eqref{eq:t_ij_t_i}.
 
The MCS procedure is a sequential testing procedure, which eliminate the worst model at each step, until the hypothesis of equal predictive ability is accepted for all the models belonging to a set of superior models. The selection of the worst model is determined by an elimination rule that is consistent with the test statistic,
\begin{equation*}
e_{\text{R,M}} = \argmax_{\rho\in M}\left\{\sup_{\xi\in M}\frac{\overline{d}_{\rho \xi}}{\sqrt{\widehat{\text{Var}}(\overline{d}_{\rho \xi})}}\right\},\qquad e_{\max, \text{M}} = \argmax_{\rho\in M}\frac{\overline{d}_{\rho.}}{\widehat{\text{Var}}(\overline{d}_{\rho.})}.
\end{equation*} 

Based on the out-of-sample MSFEs, we carried out the MCS procedure and determined the superior set of models at the 80\% and 90\% confidence levels, respectively. In Table~\ref{tab:mcs}, we consider two methods for selecting the number of retained components for the two test statistics. The RMLTS estimator gives the most accurate forecasts at both the 80\% and 90\% confidence intervals. 

\vspace{.1in}
\begin{table}[!htbp]
\tabcolsep 0.14in
\centering
\caption{MCS results based on the out-of-sample forecasts produced by FPCA, RFPCA, MLTS and RMLTS. Symbols $\dagger$ and $\ddagger$ are used to indicate that the method resides in the superior set of models at the 80\% and 90\% confidence levels, respectively.}\label{tab:mcs}
\begin{tabular}{@{}lccccccccc@{}}\toprule
& \multicolumn{4}{c}{$80\%$ confidence interval} & & \multicolumn{4}{c}{$90\%$ confidence interval} \\
Statistics & FPCA & RFPCA & MLTS & RMLTS & & FPCA & RFPCA & MLTS & RMLTS \\
\midrule
$T_{\text{R}, \text{M}}$ & & & & $\dagger$ & & & &  & $\ddagger$  \\
$T_{\max, \text{M}}$ & & & & $\dagger$ & & & & & $\ddagger$ \\
\bottomrule
\end{tabular}
\end{table}

\section{Conclusion}\label{sec:7}

The presence of outliers in functional time series is not uncommon and can have severe consequences for modeling and forecasting. We present a robust functional time series method, which first applies a robust FPCA to identify and remove outliers. Then, it decomposes a time series of curves into robust dynamic functional principal components and their associated principal component scores. Conditioning on the estimated principal components, forecast curves are obtained by accurately modeling and forecasting principal component scores. When the estimated principal component scores are modeled and forecasted via a multivariate time series method, such as the VAR models, it is advantageous to consider a robust multivariate estimator, such as the MLTS and RMLTS estimators considered here, to estimate parameters and determine the optimal lag order based on a robust information criterion. 

Via a series of simulation studies, we demonstrate the superior forecast accuracy that dynamic FPCA entails in comparison to static FPCA when there is no outlier. Furthermore, we show the superior estimation accuracy of the MLTS and RMLTS estimators compared to the OLS estimator, in the presence of outliers. Between the MLTS and RMLTS estimators, there seems to be a marginal difference in terms of point forecast accuracy. Illustrated by the hourly ozone concentration curves, the standard functional time series forecasting method is outperformed by the proposed robust functional time series forecasting method in terms of point forecast accuracy. Through the MCS procedure, the RMLTS estimator gives the most superior point forecast accuracy among the four methods considered. 

To ensure that the information presented in this paper does not diverge from the principal contribution, we will investigate the effect of outliers on interval forecast accuracy in future research. There are a number of avenues for future research. We briefly outline four below:
\begin{enumerate}
\item[1)] It is envisaged that the robust functional time series method can be extended to the functional moving average, functional autoregressive moving average (FARMA), and integrated FARMA processes. 
\item[2)] Other robust VAR estimation methods for estimating and forecasting principal component scores may be considered and compared. 
\item[3)] We present a robust method and its computational algorithm, but theoretical developments in terms of breakdown points and efficiency are worthwhile to explore, and investigation of the influential observations in a functional principal component regression is warranted. 
\item[4)] When we sequentially observe data points in the most recent curve, it is possible to apply our proposed methods to update forecasts \citep[see, e.g.,][]{Shang17}. 
\end{enumerate}
  
\newpage
\bibliographystyle{apalike}
\bibliography{robust}

\end{document}